\documentclass[twocolumn,nofootinbib,amsmath,amssymb,aps,prd,balancelastpage,superscriptaddress]{revtex4-1}

\usepackage{amsmath,amssymb}
\usepackage{amsfonts,amssymb,mathrsfs}
\usepackage{graphicx}
\usepackage[bookmarks=false,pdfstartview=FitH]{hyperref}
\usepackage[all]{hypcap}
\usepackage{epsfig,latexsym}

\RequirePackage{xspace} \allowdisplaybreaks

\def\bef{\begin{figure}}
\def\eef{\end{figure}}

\newcommand{\be}[1]{\begin{equation}\label{#1}}
\newcommand{\beq}{\begin{equation}}
\newcommand{\ee}{\end{equation}}
\newcommand{\beqn}[1]{\begin{eqnarray}\label{#1}}
\newcommand{\eeqn}{\end{eqnarray}}
\newcommand{\bd}{\begin{displaymath}}
\newcommand{\ed}{\end{displaymath}}

\def\lsim{\raise0.3ex\hbox{$\;<$\kern-0.75em\raise-1.1ex
e\hbox{$\sim\;$}}}
\def\gsim{\raise0.3ex\hbox{$\;>$\kern-0.75em\raise-1.1ex
\hbox{$\sim\;$}}}
\def\simlt{\mathrel{\lower2.5pt\vbox{\lineskip=0pt\baselineskip=0pt
           \hbox{$<$}\hbox{$\sim$}}}}
\def\simgt{\mathrel{\lower2.5pt\vbox{\lineskip=0pt\baselineskip=0pt
           \hbox{$>$}\hbox{$\sim$}}}}
\def\unity{{\hbox{1\kern-.8mm l}}}

\renewcommand{\vec}[1]{\mathbf{#1}}

\def\lsim{\mathrel{\mathop  {\hbox{\lower0.5ex\hbox{$\sim$}
\kern-0.8em\lower-0.7ex\hbox{$<$}}}}}
\def\gsim{\mathrel{\mathop  {\hbox{\lower0.5ex\hbox{$\sim$}
\kern-0.8em\lower-0.7ex\hbox{$>$}}}}}

\def\be{\begin{equation}}
\def\ee{\end{equation}}

\usepackage{color}

\begin{document}

\pagestyle{plain}

\title{Dark matter and baryogenesis in the Fermi-bounce curvaton mechanism}

\author{Andrea Addazi}
\email{andrea.addazi@lngs.infn.it}
\affiliation{Dipartimento di Fisica, Universit\`a di L'Aquila, 67010 Coppito AQ, Italy\\
 LNGS, Laboratori Nazionali del Gran Sasso, 67010 Assergi AQ, Italy}

\author{Stephon Alexander}
\email{stephon_alexander@brown.edu}
\affiliation{Department of Physics, Brown University, Providence, RI, 02912, USA}

\author{Yi-Fu Cai}
\email{yifucai@ustc.edu.cn}
\affiliation{CAS Key Laboratory for Research in Galaxies and Cosmology, Department of Astronomy, University of Science and Technology of China,\\
Chinese Academy of Sciences, Hefei, Anhui 230026, China}

\author{Antonino Marcian\`o}
\email{marciano@fudan.edu.cn}
\affiliation{Department of Physics \& Center for Field Theory and Particle Physics, Fudan University, 200433 Shanghai, China}

\begin{abstract}
\noindent
We elaborate on a toy-model of matter bounce, in which the matter content is constituted by two fermion species endowed with four fermion interaction term. We describe the curvaton mechanism that is forth generated, and then argue that one of the two fermionic species may realize baryogenesis, while the other (lighter) one is compatible with constrains on extra hot dark matter particles.
\end{abstract}

\maketitle

\section{Introduction}
\label{s.intro}

\noindent
The matter bounce scenario is an alternative to inflation that fulfills the same observational constraints as the latter, but carries definite novel predictions about CMB observables to be  measured in forthcoming experiments. At this regard the matter bounce scenario is distinguishable from inflation. Scale-invariant perturbations are generated in a contracting cosmology, which is then thought to be connected to the current phase of expansion of the universe thanks the emergence of a non-singular bounce in the dynamics. This is the theoretical peculiarity of matter bounce models with respect to the inflationary ones, as the cosmological singularity si solved, and completeness of geodesics is restored.

Cosmological perturbations are also dealt with peculiarly in each one of the frameworks. In inflation the different dynamical evolutions of the causal horizon and Hubble horizon are at the origin of the generation of scale-invariant Fourier modes that reenter the horizon. In the matter bounce is during the phase of matter-dominated contraction that Fourier modes of the co-moving curvature perturbation become scale-invariant. For a detailed introduction to the generation of scale-invariant perturbations we refer to \cite{Brandenberger:2012zb}, while for a recent review on the status of matter bounce cosmologies we refer to \cite{Brandenberger:2016vhg}.

Similarly to inflation simple realizations of the matter bounce scenario have been developed that deploy scalar matter fields, whose potentials are chosen ad hoc so to reproduce a vanishing pressure during the matter-dominated phase of contraction of the universe \cite{Cai:2013kja}. Differently than inflation, observations allow to rule out the matter bounce scenario with a single scalar field \cite{Quintin:2015rta}. Indeed single scalar field matter bounce models predict an exactly scale-invariant spectrum, while the actual observed one has a slight red tilt with a spectral index of $n_s = 0.968 \pm 0.006 ~(65\%)$ \cite{Ade:2015xua}, and a tensor-to-scalar ratio $r$ significantly larger than the value allowed by the observational bound $r < 0.12 ~(95 \%)$ \cite{Ade:2015tva}.

Nonetheless, there are few instantiations of the matter bounce scenario that predict a slight red tilt in the spectrum of scalar perturbations and fulfill the constraints on the tensor-to-scalar ratio \cite{WilsonEwing:2012pu, Cai:2014jla, Cai:2015vzv}. The mechanisms that are usually considered at this purpose hinge on the inclusion of additional matter fields \cite{Cai:2013kja, Cai:2011zx}, on the choice of a matter field that has a small sound speed (so to enhance the amplitude of vacuum fluctuations) \cite{Cai:2014jla}, and finally on the suppression of the tensor-to-scalar ratio during the bounce that is explained accounting to quantum gravity effects \cite{Wilson-Ewing:2015sfx}.

Here we will follow a different theoretical perspective, closer to the intuition developed in particle-physics. Indeed, we intend not to deploy exotic matter fields, or matter fields that have not been observed yet in terrestrial experiments, and not to resort to quantum gravity effects, extending our framework up to Planck scale. In a more conservative fashion we rather consider here matter fields that belong to the standard model (SM) of particle physics, and that correspond to the simplest and most conservative extensions of it, so encode dark matter in the picture we will develop. And following the particle-physics intuition that to a definite energy scale will correspond definite physical degrees freedom, we assume as in Ref.~\cite{Cai:2015vzv} that both the energy scale and the matter content of the universe during its contracting phase are comparable to the one of the present universe, which bring us to consider the importance of dark matter during the pre-bounce matter phase contraction of the Universe.

We wish to remark that recently bouncing cosmologies involving dark matter (and dark energy) have received much attention in the literature, and that distinctive and falsifiable  predictions on CMB observables have been derived that will be tested in the near future \cite{Cai:2015vzv, deHaro:2015wda, Odintsov:2015zua, Lehners:2015mra, Ferreira:2015iaa, Brandenberger:2016egn, Nojiri:2016ygo, Odintsov:2016tar} (for a recent review see also \cite{nuovo}). With respect to this vast literature the gist of our proposal relies on the deployment of fermionic matter fields.

Specifically, we develop here a toy-model in which both matter and dark matter are described by fermionic fields, the dynamics of which is governed by the Dirac action on curved space time, and a four fermion interaction term. The latter term is actually due to the resolution of the torsional components of the gravitational connection with respect to fermionic bilinears, and must be accounted for in the first order formalism. We then implement a curvaton mechanism, in which the fermion field with lighter mass is responsible for the generation of almost scale-invariant curvature perturbation modes, and the heavy mass field drives the dynamics of the background. We then argue that while the light fermion field can be assumed to be a neutrino, the heavy fermion field can be related to the sterile neutrino, and hence by decaying the lighter neutrinos can accommodate baryogenesis through leptogenesis.

We start in Sec. II by differentiating our approach from the many other ones present within the literature. In Sec. III we then review the instantiation of the matter bounce mechanism that deploys one fermionic field, which from now on we will call Fermi bounce cosmology. In Sec. IV we review the curvaton mechanism for a Fermi bounce cosmology that accounts for two fermionic species. In Sec. V we deepen the phenomenological consequences that can be derived for CMB observables, and comment on the falsifiability of this scenario with respect to introduction of dark matter. In Sec. VI we study the application of this curvaton model to leptogenesis, and comment on the phenomenological constraints that can be inferred from data. Finally, in Sec. VII we spell some outlooks and conclusions.

\medskip

\section{The matter bounce scenario}

\noindent
It is now days common knowledge that FLRW metrics suffer from singularities in all the curvature invariants. It was already remarked by Hawking and Penrose \cite{HE} that the initial singularity is unavoidable if space-time is described by General Relativity and matter undergoes null energy conditions (NEC).  Many non-singular bouncing cosmologies have been hitherto developed in order to solve the Big-Bang singularity issue, but at the cost of dismissing some of the assumptions behind the Hawking-Penrose theorem, most notably NEC.

Bouncing mechanisms can be implemented within frameworks very different from one another. A complete review, comprehensive of all the bouncing models developed hitherto, would be too long to be drawn in this paper, turning far away from our current purpose of focusing on a model of bounce cosmology that accounts for dark matter and only involves fermionic matter fields. Nonetheless, before focusing on fermionic matter bounce models and their instantiations able to encode dark matter, we wish to briefly survey the vast scenario offered within the literature, and enlighten some paradigmatic cases that have received much attention.

The bouncing behavior of the universe at early time can be indeed reconstructed from high-energy theory corrections to the effective equation of motion of the gravitational field. It is then worth to mention that quantum theories of gravity, as well as effective models inspired by the problem of quantum gravity, have driven many authors efforts in this sector. At this purpose, a characterization of the bouncing mechanisms inspired by loop quantum gravity and its cosmological applications --- loop quantum cosmology --- has been outlined in detailed analyses \cite{WilsonEwing:2012pu, Cai:2014zga}.

On the other side, there exists a flourishing literature that takes into account bouncing models from the point of view of string theory, for a complete review of which we refer to Refs.~\cite{Durrer:1998sv,Biswas:2005qr, Brandenberger:2013zea} as preliminary introductions. The so called Ho{\v r}ava-Lifshitz  proposal can also achieve a bouncing phase for early time cosmology, as emphasized in \cite{Brandenberger:2009yt}, while the contiguity to the bouncing scenario of $f(R)$ and Gauss-Bonnet theories can be read out respectively from Refs.~\cite{Odintsov:2014gea} and \cite{Bamba:2014mya}.

Nevertheless, the bouncing scenario does not necessarily require (quantum) gravitational corrections to the energy density, but in stead a vast literature is deploying fields that violate the null energy condition in order to achieve the bounce. Among many examples that can be pointed out, we may cite the ghost condensate scenario \cite{Lin:2010pf}, the so called Fermi bounce mechanism \cite{Alexander:2014eva, Alexander:2014uaa} and the Lee-Wick theory \cite{Cai:2008qw}. Because of their peculiarity of resulting from known theories of particle physics, which have been corroborated on the flat gravitational background by means of high-energy terrestrial experiments, we will focus in the next section on the Fermi bounce models.

\section{One field Fermi-Bounce cosmology}
\label{s.bounce}

\noindent
The action for the matter-gravity sector under scrutiny results from the sum of the gravitational Einstein-Hilbert action, further endowed with a topological term {\it \`a la} Holst, plus a non-minimal covariant Dirac action. Following previous literature~\cite{literature}, we may refer to this theory as the Einstein-Cartan-Holst-Sciama-Kibble theory (ECHSK). In the first order formalism, when gravity is coupled to fermion fields, we must allow for a torsionful part of the spin-connection. Thus the ECHSK one is necessarily a theory of gravity with torsionful connection \cite{meno1}. Nonetheless, a second order approach is always possible \cite{meno2}, which adopts a torsionless (Levi-Civita) connection $\tilde{\omega}[e]$, and thus differs from the former first order treatment in that a four fermion interaction term emerges. Notice however that the action written, no matter if cast in terms of a torsionful of Levi-Civita connection, is invariant under diffeomorphisms and local Lorentz transformations.

From now on we will focus on the ECHSK theory, which in the first order formalism reads
\begin{eqnarray}
\nonumber
{S}_{\rm Holst} =
\frac{1}{2 \kappa} \int_{M}\!\!  d^{4}x \;|e| \, e^{\mu}_{I}e^{\nu}_{J}
P^{IJ}_{\ \ \ KL}F^{\ \ KL}_{\mu \nu}(\omega)\, ,
\end{eqnarray}
in which
\begin{eqnarray} \nonumber
F^{\ \ IJ}_{\mu \nu}(\omega)=d\omega^{IJ}+\omega^{IL}\wedge\omega_L^{\ \ J}
\end{eqnarray}
is the field-strength of $\omega^{IJ}$, the Lorentz spin-connection, $\kappa = 8 \pi G_{\rm N}$ is the square of the reduced Planck length, and
\begin{eqnarray}
\nonumber
P^{IJ}_{\ \ \ KL}=\delta^{[I}_{K} \delta^{J]}_{L} - \frac{1}{2 \gamma}\epsilon^{IJ}_{\ \ KL}
\end{eqnarray}
involves the Levi-Civita symbol $\epsilon_{IJKL}$ and the Barbero--Immirzi parameter $\gamma$. The Dirac action reads $S_{\rm Dirac} = \!\frac{1}{2 }  \int d^{4}x |e| \mathcal{L}_{\rm Dirac}$, in which
\begin{eqnarray}
\nonumber
\mathcal{L}_{\rm Dirac} = \!\frac{1}{2 } \!\left[\overline{\psi}\gamma^{I}e^{\mu}_{I}\!\left(1-\frac{\imath}{\alpha}\gamma_{5}\right)\!\imath \nabla_{\mu}\psi -  m \overline{\psi} \psi \right] + {\rm h.c.}
\, ,
\end{eqnarray}
$\alpha\in \mathbb{R}$ representing the non-minimal coupling parameter. \\
The crucial observation \cite{PR, Freidel:2005sn} is that the torsionful part of the spin-connection can be integrated out of the ECHSK action via the Cartan equation (see {\it e.g.} \cite{paper00}), which is in stead recovered varying the ECHSK action with respect to the spin-connection $\omega^{IJ}$. One finally finds that the total action $S_{\rm Tot}= S_{\rm EH}+S_{\rm Dirac} +S_{\rm Int}$ in which the Einstein-Hilbert action $S_{\rm EH}$ and the Dirac action $S_{\rm Dirac}$ are cast in terms of metric compatible variables, and an additional term $S_{\rm Int}$ is present, which encodes a four fermion interaction potential. Specifically, the Einstein-Hilbert action, in terms of the metric compatible variables $\tilde{\omega}(e)^{IJ}$, reads
\be
\nonumber
S_{\rm EH}= \frac{1}{2 \kappa} \int_{M} \!\!\! d^4 x |e| e^\mu_I e^\nu_J R_{\mu\nu}^{IJ} \,,
\ee
while the Dirac action $S_{\rm Dirac}$ on curved space-time, once the covariant derivative with respect to the torsionless connection has been denoted with $\widetilde{\nabla}_\mu$, recasts as
\be
\nonumber
S_{\rm Dirac}= \frac{1}{2} \int_{M} \!\!\! d^4 x |e| \left( \
\overline{\psi} \gamma^I e^\mu_I \imath \widetilde{\nabla}_\mu \psi - m \overline{\psi} \psi \right) +{\rm h.c.}\,.
\ee
The interaction four fermion potential casts
\be
\nonumber
S_{\rm Int} \!=\! -\xi \kappa\! \int_{M} \!\!\! d^4 x |e| \,J_5^L\, J_5^M\, \eta_{LM}\,,
\ee
in which the we have used the definition of the axial current $J_5^L=\overline{\psi} \gamma_5 \gamma^L \psi$, and introduced a function $\xi$ of the real parameters $\alpha$ and $\gamma$, nameley
\be \nonumber
\xi:= \frac{3}{16} \!\frac{\gamma^2}{\gamma^2+1}\! \left(1 + \frac{2}{\alpha \gamma} -  \frac{1}{\alpha^2} \right)\,.
\ee
Notice that the sign of $\xi$ is crucial while discussing the physical applications in cosmology of the ECHSK action. For instance, a positive value of $\xi$ corresponds to a cosmological Fermi-liquids scenario in which the repulsive potential is sustaining an accelerated phase of expansion of the Universe \cite{ABC}. Conversely, a negative value of $\xi$ is providing a violation of NEC, and hence is determining a bounce in cosmological~\cite{Alexander:2014eva,  Alexander:2014uaa} or astrophysical scenarios --- for instance in Ref.~\cite{paper00} it was shown that for a suitable choice of the parameters' space region of the theory black hole may never form.

It is worth to express a comment on the most common objection against this type of models, which concerns the eventual appearance of instabilities. For instance, it has been shown in Ref.~\cite{Dubovsky:2005xd} that some scalar fields' actions that violate NEC might hold ghost and/or tachyonic instabilities, which naturally suggests similar issues might arise in the Fermi bounce context. Nonetheless, despite the analysis of Ref.~\cite{Dubovsky:2005xd} has been performed under very general assumptions, still it relies heavily on the effective Lagrangian being second or higher order in the space-time derivatives, so the same conclusions can not be easily extended to any generic action, nor to a fermionic action, which is not quadratic in the canonical momenta. Further work is needed to show whether for the latter system the linearity in the canonical momenta prevents from stability issues. For example, in context of Galileon models, several examples in which the Null energy Condition can be consistently violated, without any instabilities, was found in Refs.~\cite{Rubakov:2013kaa,Elder:2013gya,Rubakov:2014jja}.

Nonetheless, we spell here a simple argument in favor of stability that deploys mean field approximation. It is not difficult to show (for a detailed discussion, we refer the reader to \cite{AM}) that the four fermion interaction potential can be recast as a redefinition of the mass. As hint, we may consider to Fierz decompose the four fermion potential, and then focus only on the lower energy-channel of the decomposition, which entails for the densitized field $\chi=a^{3/2} \psi$, from which we can construct bilinear perturbations,
\be
\nonumber
\left( \gamma^I e_I^\mu \imath \widetilde{\nabla}_\mu - m - 2 \xi\kappa \sqrt{-g} \,\langle \overline{\chi} \chi\rangle \right)  \chi=0\,,
\ee
the brackets denoting the expectation value of the background fermion bilinear in the mean field approximation. At every energy-scale lower than the energy scale of the bounce, the effective mass will remain positive, or at most --- as it happens at the bounce --- it will vanish. This suggests there should not be any issue of instabilities that can be originated on the perturbed fields.

Any successful theory of the early universe must be able to reproduce the observed nearly scale invariant spectrum of adiabatic fluctuations in the CMBR.  Scale invariance has been investigated hitherto within the framework of bouncing models with a contracting phase, such as Ekpyrotic \cite{Ekpy}, String Gas \cite{stringas} and Pre-Big-Bang scenarios \cite{BF}.  On the other hand, for a number of these models it has proven difficult to obtain adiabatic scale invariant fluctuations in the contracting phase, mainly due to issues in resolving the singularity or mode matching between contracting and expanding phases \cite{BF}.

Nonetheless, Brandenberger and Finelli, and independently Wands \cite{BF,Wands}, have shown that a scale invariant power spectrum can be generated in a matter dominated contracting universe, proving the existence of some ``duality'' between the scale invariant power spectrum generated in the inflationary epoch and a contracting matter dominated phase.  During this latter phase, gauge invariant perturbations that cross the Hubble-scale turn out to be scale invariant because of the time-behavior of the scale factor --- in the cosmological time, $a(t) \!\sim\! (-t)^{2/3}$.  It is also worth mentioning that at non-singular bounces, scale-invariant modes are matched to scale-invariant modes in the expanding phase.

A very seminal investigation on the role of fermion fields in cosmology was reported in Ref.~\cite{AP}, in which a wide class of generic potentials of the Dirac field's scalar bilinear was considered, and a detailed scrutiny of different cosmological scenarios was made available. The first analyses of the Fermi bounce mechanism then trace back to Refs.~\cite{AB, Pop}, in which the authors realized that a torsion induced four-fermion interaction might yield a non-singular bounce. Further developments include the study of Ref.~\cite{Magueijo:2012ug}, deepening within the framework of a torsion-free theory, the role of a parity-violating four fermion self-interaction term. Production of scale-invariant scalar curvature perturbations has been finally investigated in Ref.~\cite{Alexander:2014eva}, for the case of one fermion species, and then extended in Ref.~\cite{Alexander:2014uaa} to case of the curvaton mechanism.

By considering a non-minimal coupling in the Dirac action (see {\it e.g.} Refs~\cite{M, BD, BD2}) and a topological term for the torsionful components of the spin-connection $\omega^{IJ}$ (see for instance Ref.~\cite{PR}), the inspection within Refs.~\cite{Alexander:2014eva, Alexander:2014uaa} has considerably enlarged the parameters's space of the fermionic theories previously examined in view of a bounce. This has allowed not only for a four fermion interaction which is regulated by the parameters of the theory via the $\xi$ function, but also for the emergence of a scale-invariant power-spectrum. It is indeed thanks to the presence of a torsion background that the topological term within the Holst action turns from a surface term into a contribution to the four fermion interaction term. While is this latter term that entails an almost scale-invariant power-spectrum of gravitational scalar perturbations.

Notice however that the four-fermion density modifies the Friedman equations to have a negative energy density that redshifts like $\sim a(t)^{6}$, thus the issues with anisotropies are not yet solved in this scenario, when we only take into account the tree-level contributions to the energy density. Nonetheless, quantum corrections to the effective action of fermion fields may provide an ``ekpyrotic-like'' contribution that redshifts faster that $\sim a(t)^{6}$, and is then able to wash out anisotropies when the universe approach the non-singular bounce \cite{DMG}.

Finally, we should also mention an important issue, with relevant observational consequences for the observation of power-spectra and cross-correlations function of the CMBR. This has to deal with the semiclassical limit of fermion fields, and the appropriate way of dealing with objects that fulfill the Pauli exclusion principle. Dirac fields indeed become physical observable that satisfy micro-causality only when they form bilinear that belong to the Clifford algebra. We shall then always deal with these combinations of fields, while adopting macroscopic states that represent coherent states in group theoretical meaning --- coherent fermionic states are SU(2) coherent states, known in condensed matter ad BCS states of superconductivity. We refer for this discussion to the work developed in Ref.~\cite{Dona:2016fip}.

We close this section emphasizing that the advantage of the Fermi bounce mechanism mainly relies on the fact that it does not require the existence of any fundamental scalar field not observed through terrestrial experiments in order to drive the space-time background evolution. The fermionic field added to the gravitational action is sufficient to account both for the matter bounce scenario and the generation of nearly scale-invariant scalar perturbations.

\section{Two fields Curvaton mechanism}
\label{s.bounce}

\noindent
In this section we review the curvaton mechanism for Fermi bounce cosmologies, which was first studied in Ref.~\cite{Alexander:2014uaa}. We will deploy a similar strategy than the one outlined in Ref.~\cite{Alexander:2014uaa} for the analysis of the curvature perturbations. Nonetheless, we are aware that, in order to prove cosmological perturbations of fermionic fields to be non vanishing at the linear order, the procedure first described in Ref.~\cite{Dona:2016fip} must be implemented. We will move then consistently, along the lines drawn in Ref.~\cite{Dona:2016fip}. We remark anyway that the manipulation of the perturbed fermionic bilinear that we perform here will give at the end very similar results as in Ref.~\cite{Alexander:2014uaa}. There are only few differences that concern the observable quantities for CMBR, namely the scalar power spectrum and the tensor to scala ratio parameter $r$, but this are not significant experimentally, and are totally due to the four fermion interaction between the two fermionic species that we are taking into account here.

Differently than from the approach within Ref.~\cite{Alexander:2014uaa}, in which four fermion interaction terms were added for each one of the fermionic species considered by simply following a phenomenological recipe, the four fermion terms we focus on here follow directly from the ECHSK action. As a consequence, when two fermionic species are taken into account a novel four fermion interaction term between the two species arises. We will then show that once a mass hierarchy between the two species is considered, the curvaton mechanism is again realized: the spacetime background evolution encode a bounce, and a scale invariant scalar power spectrum is generated.

\subsection{The ECHSK action and the background dynamics}

\noindent
We may start directly from the action for gravity and Dirac fermions in which torsion has been integrated out, namely
\be
S= S_{\rm GR} + S_{\rm \psi}+S_{\rm \chi}
+S_{\rm Int}\,,
\ee
where again the Einstein-Hilbert action is written using mixed-indices Riemann tensor $R_{\mu\nu}^{IJ}=F_{\mu\nu}^{IJ}[\widetilde{\omega}(e)]$, i.e.
\be
S_{GR}= \frac{1}{2 \kappa} \int_{M} \!\!\! d^4 x |e| e^\mu_I e^\nu_J R_{\mu\nu}^{IJ} \,,
\ee
the Dirac action $S_{\rm \psi}$ on curved space-time casts
\be
S_{\rm \psi}= \frac{1}{2} \int_{M} \!\!\! d^4 x |e| \left( \
\overline{\psi} \gamma^I e^\mu_I \imath \widetilde{\nabla}_\mu \psi - m_\psi \overline{\psi} \psi \right) +{\rm h.c.}\,,
\ee
and finally the interaction terms of the theory read
\be \label{interact}
S_{\rm Int} \!=\! -\xi \kappa\! \int_{M} \!\!\! d^4 x |e| \,\left( J_\psi^L\, J_\psi^M + 2J_\psi^L\, J_\chi^M +J_\chi^L\, J_\chi^M \right)\, \eta_{LM}\,,
\ee
which only involve the axial vector currents $J_\psi$ and $J_\chi$ of the $\psi$ and $\chi$ fermionic species, but in the three possible combinations. \\

For the two fermionic species the Dirac Lagrangians, provided with interactions, respectively read
\begin{eqnarray}
\mathcal{L}^{\rm Tot}_{\rm \psi} \!= &\phantom{a}&\!\frac{1}{2}\! \left( \overline{\psi} \gamma^I e^\mu_I \imath \widetilde{\nabla}_\mu \psi - m_\psi \overline{\psi} \psi \right) \!+ {\rm h.c.} \nonumber\\
&\phantom{a}& -  \xi \kappa\, J_\psi^L (J_\psi^K+J_\chi^K)\, \eta_{LK}\, ,
\end{eqnarray}
and
\begin{eqnarray}
\!\!\mathcal{L}^{\rm Tot}_{\rm \chi} =&\phantom{a}& \frac{1}{2} \left( \overline{\chi} \gamma^I e^\mu_I \imath \widetilde{\nabla}_\mu \chi - m_\chi \overline{\chi} \chi \right) \!+ {\rm h.c.} \nonumber\\
&\phantom{a}&-  \xi \kappa\, J_\chi^L (J_\chi^K+J_\psi^K)\, \eta_{LK}\,,
\end{eqnarray}
with the energy-momentum tensors
\be \label{tenfepsi}
\hspace{-0.25cm}
T^{\rm \psi}_{\mu\nu}\!=\! \frac{1}{4}  \overline{\psi} \gamma_I e^I_{( \mu} \imath \widetilde{\nabla}_{\nu )} \psi +{\rm h.c.}  -
 g_{\mu\nu} \mathcal{L}^{\rm Tot}_{\rm \psi} \,,
\ee
and
\be \label{tenfechi}
\hspace{-0.25cm}
T^{\rm \chi}_{\mu\nu}\!=\! \frac{1}{4}  \overline{\chi} \gamma_I e^I_{( \mu} \imath \widetilde{\nabla}_{\nu )} \chi +{\rm h.c.}  -
 g_{\mu\nu} \mathcal{L}^{\rm Tot}_{\rm \chi} \,.
\ee
The background dynamics of the fermionic bilinears must be solved along the lines of \cite{Dona:2016fip}. Nonetheless, conclusions are here quite similar to what was found in \cite{Alexander:2014uaa}. On the states of semiclassicality (respectively) for the $\psi$-fermion field, namely the coherent state $|\alpha_\psi\rangle$, and for the $\chi$-fermion field, namely the coherent state $|\alpha_\chi\rangle$, we can easily recover (see e.g. Ref.~\cite{Dona:2016fip}) that on shell
\be \label{psi_chi}
\langle \bar{\psi} \psi \rangle_{\alpha_\psi}= \frac{n_\psi}{a^3}\,, \qquad \langle  \bar{\chi} \chi\rangle_{\alpha_\chi}= \frac{n_\chi}{a^3}\,,
\ee
in which the fermionic densities arise from the integration of the modes' distributions of the coherent states, i.e.
\be \label{psi_chi_2}
n_\psi=\int d\mu(\vec{k}) |\alpha_\psi(\vec{k})|^2\,, \qquad n_\chi=\int d\mu(\vec{k}) |\alpha_\chi(\vec{k})|^2\,,
\ee
$d\mu(\vec{k})$ denoting the appropriate relativistic measure on the Fourier modes space.

Using the Fierz identities, an evaluating on the coherent states the prodcut of the fermionic bilenears, the first Friedmann equation can be cast, accounting for the contributions due to the two fermionic species, as it follows
\begin{eqnarray} \label{Hh}
H^2= \frac{\kappa}{3} \, \frac{m_\psi  n_\psi +m_\chi  n_\chi}{a^3}+ \xi\,\frac{\kappa^2}{3}\,\frac{(n_{\psi}+n_\chi)^2}{a^6} \,,
\end{eqnarray}
in which the double product of fermionic densities in the last term now accounts for the interaction between the two fermionic species.

The scale factor of the metric is easily determined to be
\be
\label{adri}
a\!=\!\!\left( \frac{3  \kappa (m_\psi n_\psi\!+\!m_\chi n_\chi ) }{4} (t-t_0)^2 \!-\! \frac{ \xi  \kappa \,( n_\psi + n_\chi)^2 }{(m_\psi n_\psi+m_\chi n_\chi)}  \right)^{\frac{1}{3}}\!,
\ee
and its value in $t_0$, when the bounce takes place, immediately follows
\be \label{azero}
\!\!\!a_0\!=\!\!\left( - \frac{ \xi  \kappa \,( n_\psi + n_\chi)^2 }{m_\psi n_\psi+m_\chi n_\chi}  \right)^{\frac{1}{3}} \!\simeq\! \left( - \frac{ \xi  \kappa \,( n_\psi + n_\chi)^2 }{m_\psi n_\psi}  \right)^{\frac{1}{3}} \!.
\ee

\subsection{The cosmological perturbations}

\noindent
Cosmological perturbations can be studied in the flat gauge, in which the curvature perturbation variable results to be proportional to the perturbation of the energy density of the system
\be \label{pro}
\zeta= \frac{\delta \rho}{\rho+p}\,.
\ee
Perturbations of the energy densities of the fermionic species are linear in the perturbations of the fermionic bilinear. In \cite{Dona:2016fip} it has been shown that linear perturbations of the fermionic bilinear are non-vanishing. Thus also the curvature perturbation variable defined in \eqref{pro}, linear by definition, is non-vanishing for fermion fields.

We remind that within the formalism introduced in \cite{Dona:2016fip}, given a generic operator $\mathcal{O}$ in the spinorial internal space, the $n$-th infinitesimal order expansion of the expectation value (on a quantum macroscopic coherent states) of the fermionic bilinear $\overline{\psi} \mathcal{O} \psi $ is defined by the expansion
\be
\delta^n (\langle  \overline{\psi} \mathcal{O} \psi \rangle_{ \alpha_\psi})\equiv n! \,\langle \alpha_\psi'|\overline{\psi} \mathcal{O} \psi | \alpha_\psi'\rangle \Big|_{O(\delta \alpha_\psi^n )}\,,
\ee
in which the perturbation of the modes distribution function  $\alpha_\psi'\simeq\alpha_\psi+\delta \alpha_\psi +\dots$ has been considered. For simplicity of notation, we will remove the subscript $\alpha_\psi$, and denote perturbations of fermionic bilinears on the coherent space simply as $\delta \langle  \overline{\psi} \mathcal{O} \psi \rangle$.

If we now take into account the two fermionic species with different values of the bare mass, we will find that the two main contributions to the variation of the energy densities read
\begin{eqnarray}
\delta \rho= m_\chi \delta\langle \overline{\chi} \, \chi \rangle +
m_\psi  \delta\langle \overline{\psi} \, \psi \rangle +\dots\,,
%
\end{eqnarray}
having neglected contributions suppressed by $\xi \kappa$. On the other hand, similarly to \eqref{Hh} the denominator of (\ref{pro}) becomes
\be
p+\rho=\frac{m_\chi n_\chi  + m_\psi n_\psi}{a^3} + 2 \xi \kappa \frac{(n_\chi  +  n_\psi)^2}{a^6} \,.
\ee
Not astonishingly, at the zeroth order in $\xi \kappa$ a similar result as in Ref.~\cite{Alexander:2014uaa} is obtained. For a values of $m_\chi<\!\!<m_\psi$, this reduces the expression of the curvature perturbation variable to
\be
\zeta\simeq \frac{m_\chi  \, \delta \langle \overline{\chi} \, \chi \rangle }{m_\psi \langle\overline{\psi} \psi \rangle} \,,
\ee
if it also holds that
\be \label{ansatz}
m_\psi n_\psi \gg m_\chi n_\chi \,.
\ee

Therefore, as in Ref.~\cite{Alexander:2014uaa}, we proceed to write the autocorrelation function for $\zeta(t, \vec{x})$ in the notation of \cite{Dona:2016fip} as
\be \label{P_S_general}
 \mathcal{P}_S 
  \equiv  \mathcal{P}_\zeta = \frac{m^2_\chi}{m^2_\psi} \frac{\delta^2 \langle \overline{\chi} \chi\, \overline{\chi} \chi \rangle }{4\langle\overline{\psi} \psi\rangle^2}  ~,
\ee
in which we are now assuming that:
\begin{enumerate}
\item
$m_\psi\!>\!\!>\!m_\chi$, which entails suppression at super-horizon scale of the perturbations due to the $\psi$ field --- wavenumber of perturbations of $\psi$-bilinears are more blue-shifted than perturbations of $\chi$-bilinears;
\item
cross-correlation between perturbations of the $\psi$ field and perturbations of the $\chi$ field are negligible, since they are due to an interaction involving a graviton loop, the latter being suppressed by the forth power of the Planck mass $M_p$.
\end{enumerate}

The perturbations to the $\chi$ field are then computed resorting to the same kind of assumptions outlined in Refs.~\cite{Alexander:2014eva, Alexander:2014uaa}, but following the procedure outlined in \cite{Dona:2016fip}. The analysis we report below is showing explicitly that at perturbative level fluctuations that are recovered are free of gradient and ghost instability, which often exists in other bouncing cosmologies. This refines our previous heuristic argument exposed in Sec.~III.

First, we notice that away from the bounce the scale factor reads $a(\eta) \simeq \eta^2/\eta_0^2$, with
\be \label{eta}
\eta_0=[\kappa (m_\psi n_\psi + m_\chi n_\chi) ]^{-1/2}\,,
\ee\\
which becomes, because of the requirement specified in equation (\ref{ansatz}),
\be \label{eta}
\eta_0\simeq  (\kappa m_\psi n_\psi )^{-1/2}\,.
\ee
The dynamics of the perturbations of the $\chi$-field bilinear, which differently than in \cite{Alexander:2014uaa} is now provided with a four fermion term interaction with the $\psi$ field, can be then found once the equations of motion for the field are solved
\be
\left( \gamma^I e_I^\mu \imath \widetilde{\nabla}_\mu - m_\chi -2 \xi\kappa \langle \overline{\chi} \chi\rangle+\langle\overline{\psi} \psi\rangle \right) \chi=0\,,
\ee
in which we have used the mean field approximation for the terms arising from the four fermion interactions. Upon reshuffle of the equation of motion for the $\chi$-spinor and densitization of its components we find
\be
\left( \gamma^I e_I^\mu \imath \widetilde{\nabla}_\mu - m_\chi - 2 \xi\kappa \sqrt{-g} \,(\langle\widetilde{\overline{\chi}} \widetilde{\chi}\rangle+ \langle\widetilde{\overline{\psi}} \widetilde{\psi}\rangle) \right) \widetilde{\chi}=0\,.
\ee
Deploying the background solution for the densities of the two fermionic species, the latter equations rewrites
\be \label{Diri}
\left( \imath \gamma^\mu \partial_\mu - m_\chi \, a(\eta) -\frac{2 \xi \kappa \, (n_\chi+n_\psi)}{a^2(\eta)} \right) \widetilde{\chi}=0 \, .
\ee
The appearance in \eqref{Diri} of the density numbers for both the fermionic species is peculiar of the theory under scrutiny here, derived directly from the ECHSK action, and represents a main difference with respect to the analysis in \cite{Alexander:2014uaa}, especially for what concerns CMBR phenomenology --- the other main phenomenological difference will be spelled out in the next section, and concerns baryogenesis.\\

Following the recipe implemented in Refs.~\cite{Alexander:2014eva,Alexander:2014uaa}, we can find solutions for the spinorial components of the Dirac equation (\ref{Diri}) in terms of the spinorial functions
\begin{eqnarray}
&& \tilde{f}_{\pm h}= \frac{1}{\sqrt{2}} [\tilde{u}_{L,h} (\vec{k}, \eta)+\tilde{u}_{R,h} (\vec{k}, \eta)] \,,\nonumber\\
&&  \tilde{g}_{\pm h}=  \frac{1}{\sqrt{2}} [\tilde{v}_{L,h} (\vec{k}, \eta)+\tilde{v}_{R,h} (\vec{k}, \eta)] .
\end{eqnarray}
These are recovered by rescaling densitized spinors up to $\tilde{u}=a^{3/2} u$ and $\tilde{v}=a^{3/2} v$. Densitized spinors are in turn expressed in terms of their chiral and helical components
\begin{eqnarray}
&& \hspace{-0.8cm}
\tilde{u}(t, \vec{k}) = \sum_h \tilde{u}_h(t, \vec{k})= \sum_h \left(
\begin{array}{c} \tilde{u}_{L, h} (\vec{k}, \eta) \\ \tilde{u}_{R,h} (\vec{k}, \eta)  \end{array}\right) \xi_h\,, \\
&& \hspace{-0.8cm}
\tilde{v}(t, \vec{k}) = \sum_h \tilde{v}_h(t, \vec{k})= \sum_h \left(
\begin{array}{c} \tilde{v}_{R, h} (\vec{k}, \eta) \\ \tilde{v}_{L,h} (\vec{k}, \eta)  \end{array}\right) \xi_h\,,
\end{eqnarray}
in which we have introduced the helicity $2$-eigenspinor, written in terms of the unit vector $\hat{\vec{k}}$, which is
\be
\xi_h\!=\! \frac{1}{\sqrt{2(1-h\,\hat{k}_z)}}\!\left(\begin{array}{c}  h(\hat{k}_x - \imath \hat{k}_y) \\ \imath \hat{k}_x -h\,  \hat{k}_y \end{array}\right)\!, \ \ \hat{\vec{k}}\!\cdot\! \vec{\sigma}\, \xi_h=h \,\xi_h,
\ee
$\vec{\sigma}$ standing for the Pauli matrices.

We can recast equation (\ref{Diri}) in terms of the $\tilde{f}_h$ functions:
\be \label{completa}
\tilde{f}''_{\pm h} + \omega^2(k,\eta) \tilde{f}_{\pm h}=0\,,
\ee
in which we have introduced an effective frequency, defined by
\begin{eqnarray}\label{frequency}
&&\omega^2(k,\eta) =\\
&&= k^2 \!+\! m^2_\chi a^2\!+\! \imath m_\chi a'
+ 2 \xi \kappa (n_\chi+n_\psi) \!\!\left(\frac{m_\chi}{a} \!-\! \imath \frac{a'}{a^3}\right). \nonumber
\end{eqnarray}
In this framework $\chi$ is a curvaton, and does not contribute to drive the dynamics of spacetime background. So the condition $m_\chi \! \ll \! m_\psi$ holds naturally, and the second term of the effective frequency can be then neglected. While for what concerns the third and the last term, which are imaginary, we may proceed to smooth them out, by taking the time-averaged evolution at super-Hubble scales. Thanks to this consideration, the effective frequency will turn out to depend mainly on the gradient term $k^2$ and the effective mass term $2 \xi \kappa (n_\chi+n_\psi) m_\chi/a$, in which the effect of the new interaction between the two fermionic species is evident.

Notice also that the imaginary part of the effective frequency can be suppressed in the contracting phase far away from the bounce, which ensures initial states for fermionic perturbations to be close to the vacuum fluctuations in Minkowski space.  Eventual deviations from scale invariance of the perturbations (before exiting the Hubble radius), which may arise along with the universe's contraction because of the presence of the imaginary part in \eqref{frequency} ca ne be switched off by a proper fine tuning, requiring the imaginary part to be at most of the same order of the real one. \\

Finally, we can find solutions that interpolate among two different limits:
\begin{enumerate}
\item
when the gradient term is dominant, namely at sub-Hubble scales with $|k\eta| \gg 1$, we impose the initial condition for the Fourier modes of the fermionic bilinear, which are the observable quantity, resorting to a Wentzel-Kramers-Brillouin (WKB) approximation. This yields
\begin{eqnarray}\label{tilde_f_in}
 \tilde{f}_{\pm h} \simeq \phantom{a}^4\!\!\!\! \sqrt{\frac{m_\chi}{\!\!2 k}}  e^{-ik\eta} ~.
\end{eqnarray}
This initial condition exactly coincides with the vacuum fluctuations. Second, we study the asymptotic solution to the perturbation equation in the limit of $|k\eta| \ll 1$, {\it i.e.} at super-Hubble scales. To apply the relation  $a(\eta) \simeq \eta^2/\eta_0^2$ and Eq. \eqref{eta}, one can write down the effective mass term as
\begin{eqnarray}
 -\frac{\gamma}{\eta^2} ~~ {\rm with} ~~ \gamma = -\frac{2\xi (n_\chi +n_\psi)m_\chi}{n_\psi m_\psi} ~.
\end{eqnarray}
\item
when the (other) asymptotic solution of the equation of motion has a leading term of the form
\begin{eqnarray}\label{tilde_f_out}
 \tilde{f}_{\pm h} \simeq c(k)\, \eta^{\frac{2(1+\gamma-\sqrt{1+4\gamma})}{3-\sqrt{1+4\gamma}}}~,
\end{eqnarray}
$c(k)$ being a $k$ dependent coefficient that must be determined by matching at the moment of Hubble crossing the above two asymptotic solutions \eqref{tilde_f_in} and \eqref{tilde_f_out}.
\end{enumerate}

The asymptotic solution at super-Hubble scales is therefore determined to be
\begin{eqnarray}\label{tilde_f_sol}
 \tilde{f}_{\pm h} \simeq \phantom{a}^{4}\!\!\!\!\sqrt{\frac{m_\chi}{\!\!2 k}} (k\eta)^{\frac{1+\gamma-\sqrt{1+4\gamma}}{3-\sqrt{1+4\gamma}}}\,,
\end{eqnarray}
having now normalized the expectation value of a fermionic bilinears in the asymptotic future.\\

After the phase of contraction, the universe would eventually enter the nonsingular bouncing phase by avoiding the spacetime singularity due to the help of the background fermionic condensate. During this phase, the evolution of the Hubble parameter $H$ can be approximated as a linear function of the cosmic time $t$ \cite{Cai:2012va, Cai:2013kja}. Accordingly, the scale factor behaves roughly as $a \sim \exp(t^2)$. In this case, one can read that the evolution of the nonsingular bounce can occur when the scale factor reaches the minimal value. By inserting back into the perturbation equation \eqref{completa}, the latter can be solved both numerically and semi-analytically. The procedure is similar to the analyses performed in \cite{Cai:2008qw} (see also \cite{Quintin:2015rta} for the observational constraint on the growth of primordial fluctuations during the bounce). We can learn that, for a fast bounce with the energy scale much lower than the Planck scale, the perturbations passing through the nonsingular bouncing phase would not be much affected by the background evolution. In the present study, for simplicity we assume that the perturbations were almost conserved during the nonsingular bounce and can be inherited from the contracting phase to the expanding phase, smoothly.

\subsection{The power spectrum of scalar gravitational perturbations}

\noindent
We dispose now of all the necessary ingredients to calculate the power spectrum of the primordial scalar gravitational perturbations. Indeed, if we substitute \eqref{tilde_f_sol} into \eqref{P_S_general}, we find that the power spectrum of the primordial curvature perturbations is expressed by the relation
\begin{eqnarray}\label{P_S_sol}
 \mathcal{P}_S = \frac{m_\chi^3 |\delta \alpha_\chi|^2 }{m_\psi^2 n^2_\psi} \frac{k^2}{4\pi^2 } (k\eta)^{\frac{4(1+\gamma-\sqrt{1+4\gamma})}{3-\sqrt{1+4\gamma}}} ~,
\end{eqnarray}
in which we have used relations \eqref{psi_chi}. Exact scale invariance of the power spectrum corresponds to the value $\gamma = 2$ ({\it i.e.} $\xi = -n_\psi m_\psi/[(n_\chi +n_\psi)m_\chi]$). In this case, during the matter contracting phase, the amplitude scales as $\eta^{-2}$ and the power spectrum generated in the fermion curvaton mechanism casts\begin{eqnarray}\label{P_S_matter}
 \mathcal{P}_S = \frac{m_\chi^3 \delta n_\chi }{m_\psi^2 n^2_\psi} \frac{1}{4\pi^2 a^2 \eta^2} \,
\end{eqnarray}
in which we have assumed $\delta n_\chi= |\delta \alpha_\chi|^2$ to retain only a mild, and phenomenologically negligible, dependence on $k$.

Evaluating \eqref{P_S_matter} at the end of the matter contracting phase $t_E$, when the scale factor equals the value $a_E$, enable us to recast it as
\begin{eqnarray}\label{P_S_final}
 \mathcal{P}_S = \frac{m_\chi^3 \delta n_\chi}{m_\psi^2 n^2_\psi} \frac{{H}_E^2}{16\pi^2} ~,
\end{eqnarray}
in which we have used $\eta_E = 2/{\cal H}_E=2/(a_E H_E)$. In \eqref{P_S_final} we used the notation ${\cal H}_E$ and $a_E$ respectively for the values of the comoving Hubble parameter and of the scale factor at the end of matter contracting phase, just right before the time $t_E$ at which the phase transition takes place and perturbations, before reentering the Hubble horizon, become constant.

Slight deviations of $\gamma$ from $2$ in (\ref{P_S_sol}) entails to derive phenomenologically allowed relations for the spectral index, i.e.
\begin{eqnarray}
 n_S - 1 \equiv \frac{d\ln \mathcal{P}_S}{d\ln k} \simeq -\frac{2}{3}(\gamma-2)\,,
\end{eqnarray}
in which the power spectrum appears to be red-tilted.

\section{CMBR phenomenology and Dark Matter}

\noindent
In this section, we show how hot dark matter constrains can be satisfied in a heuristic but successful fashion within the two-fermion-species toy-model we have discussed so far. But before tackling hot dark matter constrains we first focus on the phenomenological consequences of the toy-model for CMBR observables.

\subsection{Constraints for the masse from CMBR}

\noindent
We start by commenting on the production of primordial gravitational waves, and hence on the testable implications for the scalar to tensor ratio. We notice that the dynamics of primordial gravitational waves is uniquely determined by the spacetime background dynamics, and remind that their evolution decouple from other perturbation modes at linear order. Thus the derivation of the tensor perturbations power spectrum goes along Ref.~\cite{Cai:2013kja, Cai:2014xxa}, and allows to find
\begin{eqnarray}
 \!\!\mathcal{P}_T = \frac{1}{\vartheta^2}\frac{{\cal H}_E^2}{a_E^2 M_p^2} ,~~ {\rm where}~~ \vartheta = 8\pi(2q-3)(1-3q).
\end{eqnarray}
The coefficient $q$ is typically required to be less than unity, and is determined by the detailed procedure of the phase transition, since it represents a background parameter associated with the contracting phase.

If we assume that the universe is evolving through the bounce, and neglect for the moment its fermionic contraction phase, we can estimate the maximal amplitude of the Hubble rate to be of the same order of magnitude of $\frac{m_\psi}{\sqrt{\xi}}$. So at the fermionic matter-bounce phase the amplitude of the Hubble parameter can not be bigger than $\frac{m_\psi}{\sqrt{\xi}}$, which allows us to approximate it with its maximal value $|H_E| \simeq \frac{m_\psi}{\sqrt{\xi}}$, and find the corresponding power spectrum
\begin{eqnarray}
 \mathcal{P}_T \simeq \frac{1}{\vartheta^2}\frac{m_\psi^2}{|\xi| M_p^2}\,.
\end{eqnarray}
Finally the tensor to scalar ratio, which is by definition $r\equiv \mathcal{P}_T/\mathcal{P}_S$, si easily recovered to be
\begin{eqnarray}\label{r_tts}
 r = \frac{16\pi^2}{\vartheta^2}
 \frac{m_\psi^2}{m_\chi^3} \frac{n^2_\psi}{\delta n_\chi M_p^2},
\end{eqnarray}
in which the condition of scale invariance $\gamma\simeq 2$ has been assumed. This result is the same that was already derived in \cite{Alexander:2014uaa}.

Cosmological observations constraint the power spectrum of scalar perturbations to be $\mathcal{P}_S \simeq 2.2\times 10^{-9}$ \cite{Ade:2013zuv}, and set an upper bound to the detection of primordial gravitational waves, allowing for the tensor-to-scalar ratio values within the range $r < 0.12 ~(95 \%)$ \cite{Ade:2015tva}. If we neglect, as working assumption, the null hypothesis for $r$ and estimate it with its higher bound $r\sim 0.12$, we can constraint up to two parameters of the theory. Therefore, deploying the experimental value and bound respectively for $\mathcal{P}_S$ and $r$, we use equations \eqref{P_S_final} and \eqref{r_tts} to derive constraints of the masses of the fermion fields involved.

The first constraint we can derive is on the mass of the heavy species, which is the same as in \cite{Alexander:2014uaa}, i.e.
\be \label{cosmapsi}
 m_\psi^2 \lesssim 10^{-11} \, |\xi|\, M_p^2~.
\ee
Differently than \cite{Alexander:2014uaa}, if we now assume the total mass hierarchy  (\ref{ansatz}) we find that large values of $|\xi|=(n_\psi m_\psi)/ [(n_\chi+n_\psi) m_\chi]$ are favored, once we look at values of $\gamma$ that allows for a nearly scale-invariant power spectrum ($\gamma\simeq2$). Constraint (\ref{cosmapsi}) is then linking now the mass of the heavy species to the GUT scale, if we make a proper choice of $\xi\simeq 10^{4}$.

The second constraint we can derive on the masses of the two fermionic species arise from combining equations \eqref{P_S_final} and \eqref{r_tts} into
\be
 \frac{m_\psi^2}{m_\chi^3 } \frac{n_\psi^2}{\delta n_\chi M_p^2} \sim O(10^{2})~. \label{cosmachi}
\ee
This value is actually slightly different than the one reported in \cite{Alexander:2014uaa}, and can be easily achieved within this scenario, while linking the mass of the light species to the mass of the heavy one.

\subsection{Including dark matter }

\noindent
So far we have developed a fermion curvaton mechanism consistent with the latest cosmological data. The gist of this framework is in the realization of a see-saw mechanism that has phenomenological consequences in cosmology. Assuming that the fluctuation on the abundance of the light $\chi$ fermions is related to the abundance of the heavy $\psi$ fermions by
\be \label{gora}
\delta n_\chi  \simeq \frac{n^2_\psi}{m_\chi^3}  10^{-7}\,,
\ee
we can concretely realize in this framework a see-saw mechanism. The two fermionic species accounted for should indeed correspond to:
\begin{enumerate}
\item
a regular neutrino, which would correspond in this framework to the light $\chi$-fermion species, the mass of which would be then $m_\chi<10^{-3} eV$, fulfilling in this way the constraints from Big-Bang nucleosynthesis and being consistent with all the experimental data \cite{Battye:2013xqa};
\item
a sterile neutrino, which would be driving the primordial spacetime background dynamics, namely the $\psi$ species. Its mass would be at the GUT scale for a choice of the ratio that appears in (\ref{gora}). Nonetheless, even smaller values than the GUT scale would be consistent in this model for the mass of the background species $\psi$.
\end{enumerate}

We end this section with a comment on the perturbation theory in fermion fields cosmologies. For this type of cosmologies, the stress energy tensor behaves as a perfect fluid only at the background level. But if we consider perturbations, anisotropic components may appear in the stress-energy tensor. These latter would not affect our conclusions on the curvature perturbations generated in the primordial era, but may affect the propagation of primordial gravitational waves, and thus the estimate of the scalar to tensor ratio. This is a very interesting topic which deserves a further investigation, especially in light of the considerations in \cite{Dona:2016fip}.

\section{Baryogenesis in the curvaton Fermi-bounce scheme}

\noindent
In this section, we discuss a scenario in which baryogenesis is realized from the decay of the ultra-massive fermions, described by the $\psi$-species, into the SM particles, corresponding to the $\chi$-species. Thus $\psi$ is now defined as a singlet of the SM gauge group. Consequently, this ultra massive field is not protected by the electroweak symmetry, and for the Georgi's missing singlet mechanism, it should have a mass much higher than the SM vacuum expectation value (VEV) scale \cite{Georgi}.
We will not only show that the correct Baryon asymmetry is reproduced, but we also set sever limits on the bounce scale.
In Fig.1, we obtain a number density asymmetry $|n_{B-L}|$ which is compatible with the baryogenesis consistency condition $n_{B}/s \sim 10^{-10}$.

\subsection{Minimal coupling with SM leptons}

\noindent
The $\psi$-species can be then identified with the right handed (RH) neutrino, considering a see-saw type I mechanism for the left-handed (LH) neutrino mass. A Fukugita-Yanagida leptogenesis mechanism \cite{FukugitaYanagida,Buchmuller:2005eh} can be then realized, once all the Sakharov's conditions \cite{Sak} are satisfied.

A minimal Lagrangian for this instantiation of the see-saw mechanism reads
$$\mathcal{L}=y\psi l_{\alpha} \phi^{\alpha}+m_\psi \psi^{T}C^{-1} \psi+m_{\chi}\chi^{T}C^{-1} \chi$$
$$+\, \frac{\xi}{M_{Pl}^2}  (\psi\gamma_{5}\gamma_{\mu}\psi)( \chi \gamma^{\mu}\gamma_{5}\chi)+h.c.\,,$$
in which we have reinserted $M_{Pl}^2=\kappa^{-1}$. The decay channels of the heavy fermion to the SM particles are
$$\psi \rightarrow l \phi, \,\,\,\,\psi \rightarrow \bar{l}\bar{\phi}\,,$$
in which $l$ are lepton fields and $\phi$ denotes the Higgs,
$y$ is the Yukawa matrix of $\psi$ and $l$ generations.
In particular, we will assume that the number of $\psi$-generations will be more than one.
Such a Lagrangian mediates $L$-violating channels with several different $L$-number assignment for $\psi$. For instance, if $L:\psi=0$, the first Yukawa term violates the L-number of $\Delta L=1$. However, such an assignment for the lepton number of $\psi$ is problematic, since $\psi$ could also couple to quarks, destabilizing the proton. A strong fine-tuning for the coupling of $\psi$ to the quarks should be then invoked. The most elegant choice is to consider a $L$-preserving Yukawa term, while the $L$-number is violated by the Majorana mass term.
This choice coincides with $L:\psi=1$, as for RH neutrino. On the other hand, the CP violation is encoded in the complex phases of the Yukawa coupling $y$, which is a matrix in the space of leptonic generations of $\psi$.

Below we will then consider the case of a negligible four fermion interaction among the extra massive fermion and the light one. This imposes a bound on the suppression scale $\xi M_{Pl}^{-2}$.

The decay width of the heavy neutrino $\psi$ at tree level is 
\be \label{Gamma}
\Gamma_{D}=\Gamma(\psi\rightarrow \phi l)+\Gamma(\psi \rightarrow \bar{\phi} \bar{l})=\frac{1}{8\pi}y^\dagger y \, m_\psi\,,
\ee
while the annihilation cross section $\sigma(\psi \psi \rightarrow \chi\chi)$ is assumed to be subdominant, for the moment.
Once the temperature of the universe drops below the critical value $\bar{T}=m_\psi$, the heavy neutrinos cannot  follow the rapid variation of the equilibrium distribution.
The deviation from thermal equilibrium is related to a too large number density of heavy fermions, compared to the equilibrium density. However, the heavy fermion decay, and a lepton asymmetry, could be generated owing to the presence of CP-violating transitions. CP-violating transitions involves the quantum interference between the tree-level amplitude and the one-loop diagrams (vertex and self-energy contributions).
The resulting CP violation parameters can be conveniently defined as
\be \label{epsiloni}
\epsilon_{Da}\equiv \frac{\Gamma(\psi \rightarrow l\phi)-\Gamma(\psi \rightarrow \bar{l}\bar{\phi})}{\Gamma(\psi \rightarrow l\phi)+\Gamma(\psi \rightarrow \bar{l}\bar{\phi})}\,.
\ee
At tree-level, this parameter is equal to zero.
However,  the CP asymmetry appears at the leading order of perturbation theory $y^{2}$ (1-loop), entailing the two contributions
\be \label{wave}
\epsilon_{a}^{M}=-\frac{1}{8\pi}\frac{ {\rm Im}(y^{\dagger}y)^{2}_{ik}}{(y^{\dagger}y)},\,\,\, (\rm 3level)+(\rm self\, energy)\,,
\ee
where $i,k$ are $\psi$-generation indices,
\be \label{vertex1loop}
\epsilon_{a}^{V}=-\frac{1}{8\pi} f\left(\frac{m_{\psi_{i}}}{m_{\psi_{i}}} \right)\frac{ {\rm Im}(y^{\dagger}y)^{2}_{ik}}{(y^{\dagger}y)}\,,
\ee
and
\be \label{fx}
f(x)=\sqrt{x}\left\{1-(1+x)\rm ln\left(\frac{1+x}{x} \right)\right\}
\ee
These results hold for $m_\psi>>|\Gamma|$ -for small mass differences, we may expect an enhancement of the mixing contribution.

\subsection{The Boltzmann equations }

\noindent
The generation of a baryon/lepton asymmetry can be treated with Boltzmann equations. 
While accounting for these latter, we have to consider the main processes, which are the decays and inverse decays of the heavy fermions, and the lepton number conserving $\Delta L=0$ and violating $\Delta L=2$ processes. These processes read, respectively,
\begin{eqnarray} \label{B1}
\frac{d n_{\psi}}{dt}+3Hn_{\psi}=&&-\gamma(\psi\rightarrow l\phi)+\gamma(l\phi\rightarrow \psi) \\
&&-\gamma(\psi \rightarrow \bar{l}\bar{\phi})-\gamma(\bar{l}\bar{\phi}\rightarrow \psi)\,, \nonumber
\end{eqnarray}
\begin{eqnarray}\label{B2}
\frac{d n_{l}}{dt}+3Hn_{l}=&&\gamma(\psi \rightarrow l\phi)-\gamma(l\phi \rightarrow \psi) \\
&&+\gamma(\bar{l}\bar{\phi}\rightarrow l\phi)-\gamma(l\phi \rightarrow \bar{l}\bar{\phi})\,,\nonumber
\end{eqnarray}
\begin{eqnarray} \label{lbar}
\frac{dn_{\bar{l}}}{dt}+3Hn_{\bar{l}}=&&\gamma(\psi\rightarrow \bar{l}\bar{\phi})-\gamma(\bar{l}\bar{\phi}\rightarrow \psi) \\
&&+\gamma(l\phi \rightarrow \bar{l}\bar{\phi})-\gamma(\bar{l}\bar{\phi}\rightarrow l\phi)\,, \nonumber
\end{eqnarray}
with the reaction densities expressed as it follows
\begin{eqnarray} \label{dens1}
\gamma(\psi \rightarrow l\phi)=\int d\Phi_{123}f_{N}(p_{1})|\mathcal{M}(\psi\rightarrow l\phi)|^{2},
\end{eqnarray}
and
\begin{eqnarray} \label{dens2}
\!\!\!\!\!\!\!\!\!\gamma(l\phi\rightarrow \bar{l}\bar{\phi})=\int d\Phi_{1234}f_{l}(p_{1})f_{\phi}(p_{2})|\mathcal{M}'(\psi\rightarrow l\phi)|^{2}\,,
\end{eqnarray}
where $H$ is the Hubble parameter, $d\Phi_{1,..,n}$ denotes the phase space integration over particles in the initial and final states,
\be \label{spacephase}
d\Phi_{1,..,n}=\frac{d^{3}p_{1}}{(2\pi)^{3}2E_{1}}...\frac{d^{3}p_{n}}{(2\pi)^{3}2E_{n}}(2\pi)^{4}\delta^{4}(p_{1}+...-p_{n})
\ee
and the weights
\be \label{fini}
f_{i}(p)=\rm exp(-\beta E_{i}(p)),\ \ \ \ \ n_{i}(p)=g_{i}\int \frac{d^{3}p}{(2\pi)^{3}}f_{i}(p)
\ee
represent the Boltzmann distribution -for simplicity we use Boltzmann distribution rather than Bose-Einstein and Fermi-Dirac distributions, neglecting the distribution functions in the final state, which is a good approximation for small number densities) and the number density of particle $i=N,l,\phi$ at temperature $T=\frac{1}{\beta}$, respectively, $\mathcal{M}$ and $\mathcal{M}'$ denote the scattering matrix elements of the indicated processes at $T=0$ (the prime indicated that in $2 \rightarrow 2$ scatterings the contribution of the internal resonance state has been subtracted).

The ratio of number density and entropy density (namely $Y_{X}=n_{X}/s$) remains constant for an expanding universe in thermal equilibrium. The heavy fermions are weakly coupled to the thermal bath. So that they freeze-out of the thermal equilibrium at $\bar{T}=m_\psi$.
This implies that the decay rate is too slow to follow the rapidly decreasing equilibrium distribution $f_{N}\sim \rm \exp(-\beta m_\psi)$. As a consequence, the system will evolve toward an excess of the number density $n_{N}>n_{N}^{eq}$.

As is known, the Boltzmann equations are classical dynamical equations. 
However, they 
are endowed with S-matrix elements, in turn containing quantum mechanical interferences of different amplitudes. 
The S-matrices are contained in collisions terms in the Boltzmann equations. 

The scattering matrix elements are evaluated at zero temperature.
However, the quantum mechanical interferences must be affected by interactions with the thermal bath. For $2\rightarrow 2$ scatterings, one finds 
\be \label{lphi}
|\mathcal{M}(l\phi\rightarrow \bar{l}\bar{\phi})|^{2}=|\mathcal{M}'(l\phi\rightarrow \bar{l}\bar{\phi})|^{2}+|\mathcal{M}_{res}(l\phi\rightarrow \bar{l}\bar{\phi})|^{2}\,,
\ee
where the resonance contribution reads
\be \label{resonanceco}
\mathcal{M}(l\phi\rightarrow \bar{l}\bar{\phi}) \sim \mathcal{M}(l\phi \rightarrow \psi)\mathcal{M}(\psi\rightarrow \bar{l}\bar{\phi})^{*}=|\mathcal{M}(l\phi \rightarrow \psi)|^{2}\,.
\ee
The particles involved in the process may be treated as massless ones: their distribution functions will coincide with the equilibrium distribution. On the other hand, resonances may be treated as on-shell particles, falling out of thermal equilibrium.  Because of CPT invariance, for thermal equilibrium distributions we must have
\be \label{Bhave}
\frac{d(n_{l}-n_{\bar{l}})}{dt}+3H(n_{l}-n_{\bar{l}})=\Delta \gamma^{eq}=0\,.
\ee

The resonance contributions (the decay and inverse decay)
\be \nonumber
\Delta \gamma^{eq}_{res}=-2 \epsilon \gamma^{eq}(\psi\rightarrow l\phi)\,,
\ee
have a compensating term from $2\rightarrow 2$ contribution processes
\begin{eqnarray}
&&\Delta \gamma_{2\rightarrow 2}=2 \int d\Phi_{1234}f_{l}^{eq}(p_{1})f_{\phi}^{eq}(p_{2})\nonumber\\
&&\times (|\mathcal{M}'(l\phi \rightarrow \bar{l}\bar{\phi})|^{2}-|\mathcal{M}'(\bar{l}\bar{\phi}\rightarrow l\phi)|^{2})\,,
\end{eqnarray}
Such a $l\phi \leftrightarrow \bar{l}\bar{\phi}$ compensation is a consequence of  CPT symmetry/unitarity.
From unitarity, one can find 
\be \nonumber
\sum_{X}(|\mathcal{M}(l\phi \rightarrow X)|^{2}-|\mathcal{M}(X \rightarrow l\phi)|^{2})=0\,,
\ee
where $X$ denotes all possible generic channels. In the case of weak coupling $y$, the channel is restricted to 2-particles states to the leading order of perturbation expansion. Therefore, at the leading order $y^{2}$, one obtains 
\begin{eqnarray} \label{22}
\Delta_{2\rightarrow 2}^{eq}=&&2\int d\Phi_{1234}f_{l}^{eq}(p_{1})f_{\phi}^{eq}(p_{2}) \\
&&\times \Big(-|\mathcal{M}(l\phi \rightarrow \psi)|^{2}|\mathcal{M}(\psi \rightarrow \bar{l}\bar{\phi})|^{2} \nonumber\\
&&+|\mathcal{M}(\bar{l}\bar{\phi}\rightarrow \psi)|^{2}|\mathcal{M}(\psi \rightarrow  l\phi)|^{2}\Big)\frac{\pi}{m_\psi\Gamma}\delta(s-m_\psi^{2}) \nonumber \\
&&=2\epsilon \gamma^{eq}(\psi \rightarrow l\phi)=-\Delta \gamma_{res}^{eq}\,. \nonumber
\end{eqnarray}
The incorporation of off-shell effects requires a formalism that goes beyond the Boltzmann equations (Kadanoff-Baym equations \cite{KB}). However, the corrections are expected to be negligible from Boltzmann's model. 

The numerical integration of the Boltzmann equations for reasonable parameters is displayed in Fig.1. In particular, the number density of $\psi$-particle decreases in cosmological time, generating a small Lepton number asymmetry.

\begin{figure}[t]
\centerline{ \includegraphics [height=6cm,width=1.0\columnwidth]{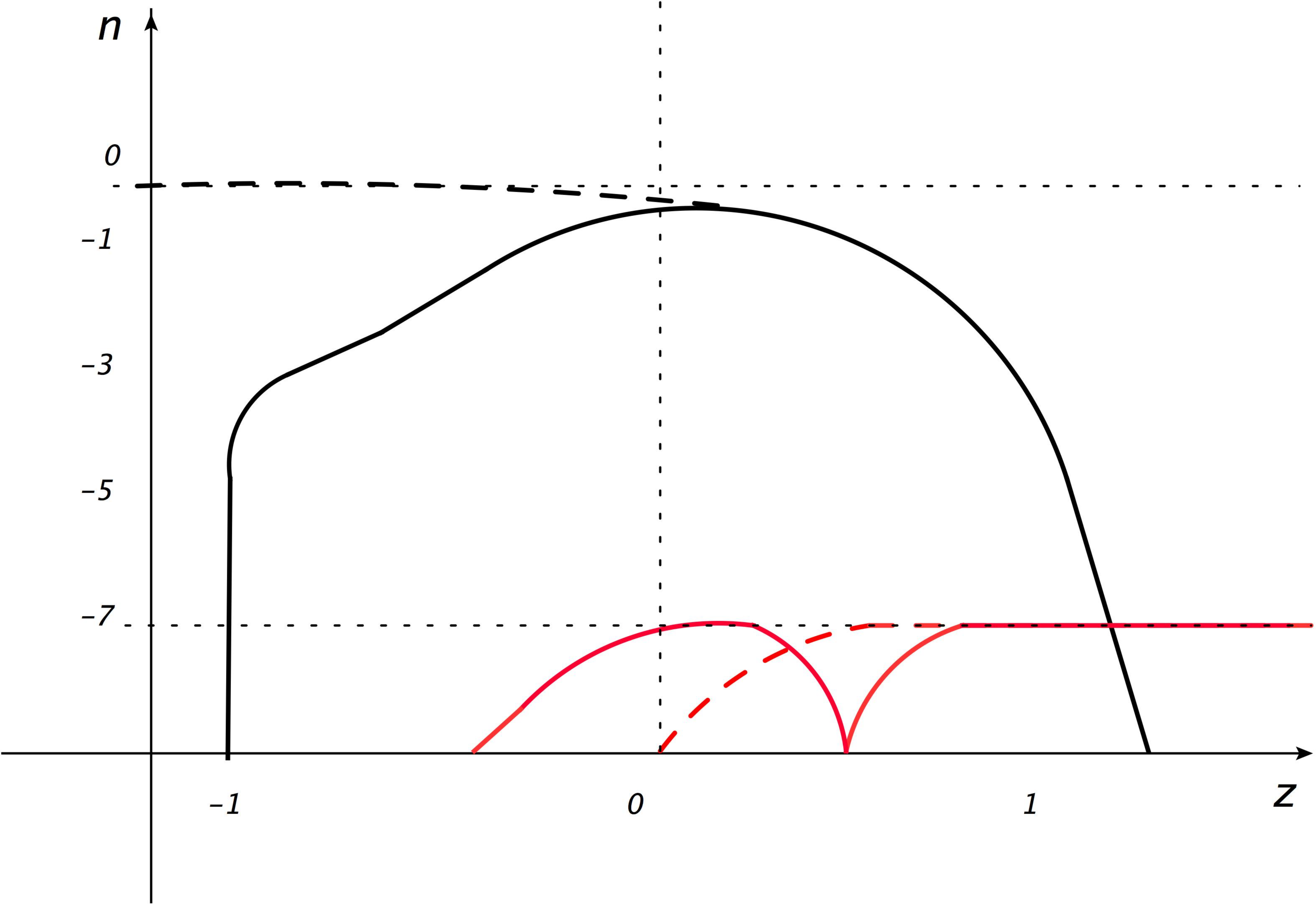}}
\vspace*{-1ex}
\caption{The evolution of the number densities as functions of $z=m_\psi/T$ are displayed $(\log_{10}n,\log_{10}z)$ scale. In dashed black lines, we show the evolution of the number density of massive fermions $\psi$ starting from a thermal initial density $n_{\psi}(z\!<\!\!<\!1)\simeq 3/4$. In thick black lines, we plot the evolution of the number density of massive fermions $\psi$ starting from a thermal initial density $n_{\psi}(z\!<\!\!<\!1)\simeq 0$. In the red dashed and thick red lines the $B-L$ number density $|n_{B-L}|$ is pictured from initial thermal and zero $n_{\psi}$ respectively. The characteristic parameters chosen in this plot are $m_\psi \simeq 10^{10}\, {\rm GeV}$, $\epsilon \simeq 10^{-6}$, $m_\chi\equiv m_{\nu_{1}}\simeq 10^{-3}\, {\rm eV}$, $\xi \leq 10^{-6}\div 10^{-8}$.}
\label{plot}   
\end{figure}

\subsection{Baryon number asymmetry from sphalerons }

\noindent
Let us assign a chemical potential $\mu$ to each SM particles: quarks, Higgs, and leptons.
In the SM -with $N_{f}$ generations and one Higgs doublet - we must assign\footnote{Let us note that, in addition to the Higgs doublet, the two left-handed doublets $q_{i}$ and $l_{i}$ and the three right-handed singlets $u_{i}$, $d_{i}$ and $e_{i}$ of each generation, will be treated with independent chemical potential (each one).} $5N_{f}+1$ chemical potentials.
The asymmetry in the particle and antiparticle number densities, for $\beta \mu_{i}<<1$, reads
\be \label{nig}
n_{i}-\bar{n}_{i}=\frac{gT^{3}}{6}\beta\mu_{i}+\mathcal{O}[(\beta\mu_{i})^{3}]\ \ \ \ \ \rm (for \ \ fermions)\,,
\ee
and
\be \nonumber
\!\!\!\!\!\!\!n_{i}-\bar{n}_{i}=\frac{gT^{3}}{6}2\beta \mu_{i}+\mathcal{O}[(\beta\mu_{i})^{3}]\ \ \ \ \ \rm (for \ \  bosons)\,.
\ee
having considered the thermal bath as a non-interacting gas of massless particles. 

However, the plasma of the early Universe is very different from a weakly coupled relativistic gas.
This because of unscreened non-abelian gauge interactions which have very important nonperturbative effects
to take into account. 

Leptons, Quarks and Higgs will interact via perturbative operators Yukawa and gauge couplings.
However, they will also interact via the nonperturbative sphaleron processes. These processes 
lead to a set of 
 constraints between the SM chemical potentials. The effective interactions induced by sphalerons ($SU(2)$ electro-weak instantons) must imply
\be \label{1CO}
\sum_{i}(3\mu_{qi}+\mu_{ui}-\mu_{di})=0\,.
\ee
On the other hand, $SU(3)$ QCD instanton-mediated processes must generate effective interactions between LH and RH quarks, which leads to constrains 
\be \label{2CO}
\sum_{i}(2\mu_{qi}-\mu_{ui}-\mu_{di})=0\,.
\ee
A third condition (valid in all the range of temperatures) arises from the requirement that the total hypercharge of the plasma must vanish. 
So that, from (\ref{nig}) and the SM hypercharges we
obtain
\be \label{hyper}
\sum_{i}(\mu_{qi}+2\mu_{ui}-\mu_{di}-\mu_{li}-\mu_{ei}+\frac{2}{N_{f}}\mu_{\phi})=0\,.
\ee
Finally, from the Yukawa couplings, one finds, in turn
\be \label{yu123}
\mu_{qi}-\mu_{\phi}-\mu_{dj}=0,\,\,\, \mu_{qi}+\mu_{\phi}-\mu_{uj}=0,\,\,\,\mu_{li}-\mu_{\phi}-\mu_{ej}=0\,.
\ee
These relations are valid if and only if the system is in thermal equilibrium\footnote{Let us note that Yukawa interactions are in equilibrium only within a {\it more restricted} temperature window depending on the strength/magnitude of the Yukawa couplings. In the following discussions, we will ignore these technical complications. 
In fact,  only a small effect on our discussion of Leptogenesis is expected from taking them into account.}, at temperatures $100\,\rm GeV<T<10^{12}\,\rm GeV$.

Let us define the Baryon number density $n_{B}=gBT^{2}/6$ and the lepton number densities $n_{li}=L_{i}gT^{2}/6$.
So that, 
 we express Baryon and Lepton numbers $B$ and $L_{i}$ in terms of the chemical potentials, i.e.
\be \label{Bden}
B=\sum_{i}(2\mu_{qi}+\mu_{ui}+\mu_{di})\,,
\ee
$$L_{i}=2\mu_{li}+\mu_{ei},\,\,\,L=\sum_{i}L_{i}\,.$$
Solving this simple system of algebric equations, with respect to $\mu_{l}$, one obtains 
\be \label{BL}
B=-\frac{4N_{f}}{3}\mu_{l},\,\,\,L=\frac{14N_{f}^{2}+9N_{f}}{6N_{f}+3}\mu_{l}\,,
\ee
They yield to the important connections between the B, B-L and L asymmetries: 
\be \label{BBL}
B=a(B-L);\,\,\,L=(a-1)(B-L)\,,
\ee
where $a=(8N_{f}+4)/(22N_{f}+13)$.

\subsection{Four fermion interaction between the two species}

\noindent
We wish now to comment on the possible relevance of the torsion mediated four fermion interactions, namely
\be
O_{\Gamma}=\frac{\xi}{M_{Pl}^{2}}\, \overline{\psi} \gamma_{\mu}\gamma_{5}\psi \  \overline{\chi} \gamma^{\mu}\gamma_{5}\chi\,.
\ee
If $\chi$ is weakly interacting with the SM particles, $\psi \psi \rightarrow \chi\chi$ represent entropy leaking collisions that could affect the leptogenesis scenario if $m_\psi\sim \xi^{-1} M_{Pl}$.  In particular, if the fermion $\psi$ has a lepton number assignment $L=0$, the $O_{\Gamma}$ operator violates the lepton number as $\Delta L=2$.  For example, the initial number density of $n_{\psi}$ can be affected by the annihilation process
\be\nonumber
\sigma(\psi\psi \rightarrow \chi \chi)\sim \frac{\xi s}{8\pi M_{Pl}^{2}}\,,
\ee
estimated\footnote{We notice that for $s>\xi M_{Pl}^{2}$ the ultra-violet (UV) completion of such a term is not yet understood. However, our analysis concerns energies which are supposed to be lower than UV cutoff scale, where unitarity issues are not expected.} at $s<\xi M_{Pl}^{2}$. This process becomes out-of-equilibrium for $s\simeq T^{2}\leq 4m_\psi^{2}$, which means that $\chi \chi$ collisions can not reproduce two heavy $\psi$ particles.

In the scenario we deepened here, we imposed the hierarchy $m_\psi<\chi^{-1/2}M_{Pl}$, i.e. the bounce scale must be higher than the heavy fermion mass. On the other hand, a successful leptogenesis requests a fermion mass scale of $m_\psi\sim 10^{9}\, \rm GeV$ or similar. This imposes an indirect bound on the bounce scale. In particular, from numerical calculations, we checked that the safety bound for a good leptogenesis is $\xi^{-1/2} M_{Pl}\geq  (10\div15) m_\psi$, assuming natural initial conditions $n_{\psi}(z<10^{-1})\simeq 0, 3/4)$ respectively. From these values, the same plots displayed in Fig.1 are obtained.

On the other hand, the case in which
$m_\psi \geq (10\div 15)\xi^{-1/2}M_{Pl}$ cannot correspond to a successful leptogenesis, without an unnatural initial superabundance of heavy fermions. In particular, the case of $m_\psi$ $\simeq$ $\xi^{-1/2}M_{Pl}$ is undesired, first of all because the unitarity and the calculability of $z<1$ in Fig.1 can not be controlled, and second because the annihilation process will dominate over all other channels
around $z\simeq 1$, provoking a too fast number density decay of $\psi$.

\section{Discussion}
\label{s.disc}

\noindent
In this paper we have shown that the matter bounce scenario, as an alternative framework to inflation, allows to encode dark matter in a pretty natural way when fermion matter fields are taken into account. We have further shown that the model is able to generate leptogenesis, thus to explain baryogenesis. Specifically, we have focused on a toy-model for the curvaton mechanism, which is an instantiation of a Fermi-bounce cosmology that singles out as a dark matter candidate a sterile neutrino-like field, hence deriving phenomenological consequences of these assumptions. It is quite remarkable that this scenario comes out to be falsifiable, since it predicts a non vanishing value of $r$.

A peculiarity of the model we have shown here is that the results we have derived are not confined to an effective analysis in which the dynamics of dark matter is only addressed from a hydrodynamical perspective. Conversely, the Fermi matter bounce scenario described here encodes microphysical description in terms of fermionic fields, which provides a natural way to overcome shortcomings that usually arise in cosmology because of the use of auxiliary scalar fields, with consequent issues of arbitrariness as it happens in inflation.

It might be indeed look surprising at a first sight that matter fields belonging to the SM and to its simplest extension reproduce desired background and perturbation features. But it should be not surprising that in this framework retrieved from particle physics it is possible to tackle questions that concern the nature of dark matter and the origin of baryogenesis. Indeed the model we have deepened here turns out to be compatible with hot dark matter constrains. Several issues continue to be unexplored, and we can not deny that this line of research will require in the future more detailed investigations. Nonetheless we wish to mention that the Fermi bounce scenario might entail as distinctive phenomenological predictions, able to falsify this paradigm among the others in the literature, the appearance of non-vanishing cross-correlation functions between polarization modes, sourced by the parity-violating elements of the Clifford algebra bilinears \cite{Dona:2016fip}. But more work is still also required even in this direction.

\acknowledgments

The work of AA~was supported in part by the MIUR research grant Theoretical Astroparticle Physics PRIN 2012CPPYP7 and by SdC Progetto speciale Multiasse La Societ\`a della Conoscenza in Abruzzo PO FSE Abruzzo 2007-2013. The work of YFC~is supported in part by the Chinese National Youth Thousand Talents Program (Grant No. KJ2030220006), by the USTC start-up funding (Grant No. KY2030000049), by the NSFC (Grant No. 11421303 and No. 11653002), and by the Fund for Fostering Talents in Basic Science of the NSFC (Grant No. J1310021). AM~wishes to acknowledge support by the Shanghai Municipality, through the grant No. KBH1512299, and by Fudan University, through the grant No. JJH1512105.

\end{document}